\documentclass[review]{elsarticle}

\usepackage{lineno,hyperref}
\modulolinenumbers[5]

\journal{Physica D}

\usepackage{graphicx}
\usepackage{caption}
\usepackage{subcaption}

\usepackage[a4paper]{geometry}
\usepackage{amsmath,amsfonts,amssymb}
\usepackage{xcolor,ulem}

\long\def\symbolfootnote[#1]#2{\begingroup%
\def\thefootnote{\fnsymbol{footnote}}\footnote[#1]{#2}\endgroup}

\newcommand{\e}{{\mathrm e}}

\begin{document}

\begin{frontmatter}

\title{Mechanics and polarity in cell motility}
\author{D. Ambrosi, A. Zanzottera}
\address{MOX-Dipartimento di Matematica, Politecnico di Milano, \\
         and Fondazione CEN, Piazza Leonardo da Vinci 32, 20133 Milano, Italy
            }
\date{\today}


\begin{abstract}
The motility of a fish keratocyte on a flat substrate exhibits two distinct regimes:
the non-migrating and the migrating one. In both configurations the shape is fixed
in time and, when the cell is moving, the velocity is constant in magnitude and direction.
Transition from a stable configuration to the other one can be produced by a mechanical or
chemotactic perturbation.
In order to point out the mechanical nature of such a bistable behaviour, we focus
on the actin dynamics inside the cell using a minimal mathematical model.
While the protein diffusion, recruitment and segregation govern the polarization process,
we show that the free actin mass balance, driven by diffusion, and the polymerized actin
retrograde flow, regulated by the active stress, are sufficient ingredients to account
for the motile bistability.
The length and velocity of the cell are predicted on the basis of the parameters of
the substrate and of the cell itself.  The key physical ingredient of the theory is
the exchange among actin phases at the edges of the cell, that plays a central role
both in kinematics and in dynamics.
\end{abstract}

\end{frontmatter}

\linenumbers

\section*{Introduction}
The motility of eukariotic cells on flat substrates is conventionally described according to four
phases: leading edge protrusion, adhesion to the substrate, contraction at the trailing edge
and then retraction of the tail \cite{alberts}. This scenario is however only stereotypical because
for some cells, like fish keratocytes, these phases follow each other so rapidly
that they are actually undistinguishable: all the steps occur simultaneously and
the translation process is continuous in time.
Moreover fish keratocytes exhibit two distinct motile regimes: in absence of external stimuli,
they typically stay at rest with a rounded shape. A sufficiently large mechanical or chemotactical signal
can trigger a destabilization that, in about 200 seconds, yields the cell to travel
with constant velocity (up to 1 micron per second) and shape \cite{verkhovsky}.
The same kinematics even characterizes the motion of a fragment of cell lamellipodium, with a lower speed
($\simeq 2-10$ micrometers per minute); as it lacks nucleus,
microtubules and most organelles, the experiment suggests that the key elements of cell crawling
are well represented even in such a simple system. The exhibited bistable behavior, signature
of a nonlinear dynamics, is a fascinating challenge for the mathematical modelling
of an active living mechanical system.

Both when the cell (or the lamellipodial fragment) is apparently at rest and when it is steadily migrating,
the inner equilibrium of the mechanical system is actually dynamical: at a subcell level, the flow
of free and polymerized actin follow patterns that have been unravelled during
the last decades \cite{mogilner09,rubinstein}. Oligomeric actin (G--actin) freely diffuses
in the cytosol and attaches to the barbed ends of the polimeric phase network (F--actin)
that point outward the cell membrane; polymerized actin is backward transported by the myosin motors
(retrograde flow). The vectorial sum of material velocity
and growth velocity (polymerization rate) at the cell boundary produces the visible speed of the cell
(which is possibly null). Free actin monomers detach from the actin network in the body of the cell and are
passively transported by Brownian motion from regions of higher to regions of smaller concentration
(namely the cell periphery, where the polymerization process acts as a sink for the G-actin) \cite{mogilner96}.
The free actin density correlates well with the high stress regions inside the cell, inhibition
of the stress due to myosin activity slows down the cell, thus suggesting that
the mechanical stress drives the depolymerization process \cite{fournier, wilson}.

The shape of the boundary and the concentration patterns are very different between crawling
and non crawling fragments.  The fragment at rest is rounded, the lamellipodium has a symmetric shape,
the actin cytoskeleton near the membrane grows at a constant rate and is backtransported from the boundary
to the interior at the same velocity it is produced, so that the vectorial sum of velocities is zero.
In a travelling fragment, the lamellipodium takes instead a characteristic
canoe shape, the symmetry of actin and myosin concentration patterns is broken and the region
of strong actomyosin activity is concentrated in the rear of the lamellipodium.
\begin{figure}[h!]
\centering
   \includegraphics[width=14cm]{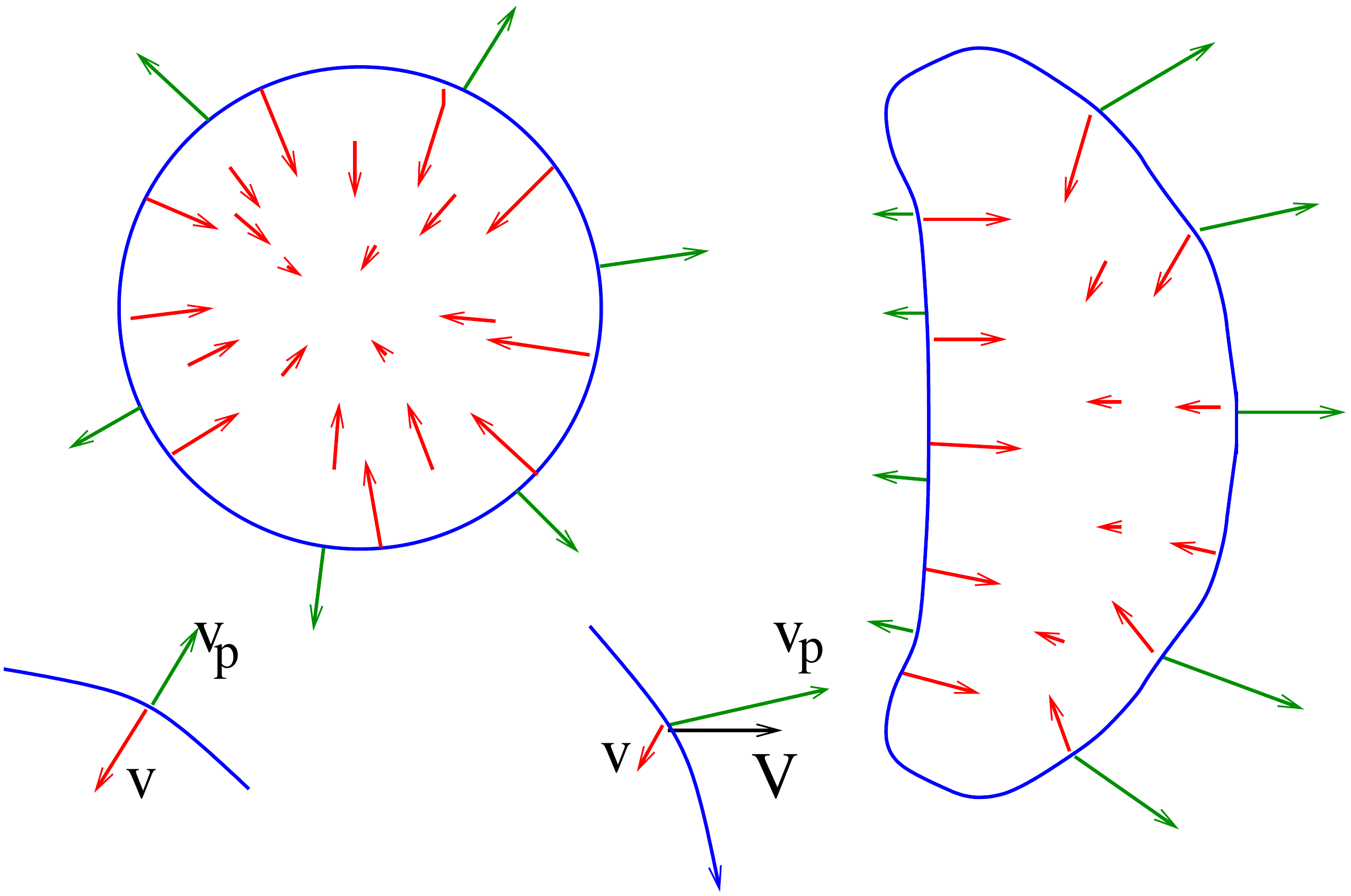}
   \caption{Velocity pattern in a fish keratocyte at rest (left) and in a migrating one (right)
            (sketch from figure 1 in \cite{wilson}).
            In the cell at rest, the material velocity $v$ (in red) balances the growth velocity $v_p$
            (in green)
            at the boundary; in a migrating cell the sum of the contributions imbalance at the boundary and
            gives rise to the stedy motion with velocity $V$ which is the vectorial sum of the
            two contributions \cite{givli}.}
\label{fig2d}
\end{figure}

The polarization of the cell and the initiation of motility is strongly linked to the migration of
several proteins on the plasma membrane and their spatial segregation. In particular,
Rho GTPases are of crucial importance in regulating actin dynamics at the cell periphery \cite{ridley2006}.
Actin nucleation is typically the rate-limiting step in actin polymerization in vivo, so cells
regulate where and when polymerization occurs by controlling the localization and activation
of actin-nucleation promoting factors \cite{keren2011}. The polarity of the network is therefore
instructed by the polarization of the membrane.

The mathematical modelling of cell migration has received an increasing attention over the last years.
The evidence of a bistable behaviour is the signature of a nonlinearity that has been encoded in
the models in terms of nonlinear constitutive equations for the cellular material.
Alt and Dembo \cite{alt} and Kimpton et al \cite{kimpton} represent a cell as made of biphasic material,
where the stress of the network has a spherical component which is cubic in the volume fraction.
The nonlinearity of the constitutive equations determine the multiplicity of the stable regimes.
According to Giomi and De Simone \cite{giomi} a cell described as an active nematic droplet
undergoes spontaneous division and motility thanks the phase separation favoured by
the Ginzburg--Landau energy density and the nonlinearity of the Landau-deGennes free energy density.
The theoretical aspects of cell motility can be addressed after homogenization of equations that
detail the dynamics of the actin network, each filament being treated as an inextensible rod
that resists to bending.
Branching, capping, cross links and friction of the filaments of the network can be very naturally included
in such a modelling framework \cite{oelz}. Other approaches focus on the contraction-driven motility
in a 1D cell: recent theoretical works based on the theory of active gels show that crawling is possible
even without polymerization \cite{recho13}.
This particular mechanism of cell motility allows a precise formulation of the condition
of optimal trade-off between performance and metabolic cost, so that
the distribution of contractile elements can be recovered on the basis of an optimization argument
\cite{recho14}.
The region at the leading edge, where the branched F-actin network is not stabilized by
cross links, is very narrow when compared with is the typical length of the lamellipodium:
a few hundreds nanometers vs 10 micrometers \cite{zimmermann2010}. An accurate mathematical
modelling of such a flexible region is needed to get rid of the force-velocity relationship
observed when a cell migrates on a flat substrate pushing a bending cantilever \cite{zimmermann2012}.

The mathematical model illustrated in this paper is inspired by Mori et al \cite{mori} and
Larripa and Mogilner \cite{larripa, novak}, where the protein recruitment and segregation
in the cytosol and on the membrane, the mechanics of the cytoskeleton and the treadmilling
of actin are separately addressed. Here
we formulate a minimal 1D theoretical model where the observed dynamics of free and polymerized actin
in the cell is coupled with the stress actively produced by the myosin motors. The polarization
of the cell follows the ``wave pinning'' model; the linear diffusion of proteins (namely Rho GTPases)
on the plasma membrane and in cytosol and the nonlinear exchange between active (membrane-bound)
and inactive (cytosolic) forms explains the existence of polar and non-polar
stable states, while the (reversible) transitions among such states of the system occurs by
finite perturbations  \cite{mori}.  The formation of a stationary front (the ``wave pinning''),
sharply separating the cell into two regions, mechanically identifies
the polarity of the F-actin network, where barbed ends do not point out of the cell
in a radially symmetric way, but mostly in the direction of the motion.
The novelty of our work is the focus on the mass conservation and momentum balance laws of the
actin phases, supplemented by mass exchange and boundary flow, that all the components
of the system must satisfy. While bistability is governed by diffusion and exchanges of proteins,
we mechanically enforce the polarity of the cell in the boundary condition for the F-actin,
relating its concentration to local amount of the active proteins.
The linearity of the constitutive equations allows to analytically compute stress
and concentration fields, that are qualitatively in agreement with the observations.
Our main result is the quantitative prediction of the length and the migration velocity
of the cell fragment in the two stable configurations (cell at rest and motile cell)
as a function of the physical parameters of the model.

\section{The mathematical model}
\label{s1}
The lamellipodium of a fish keratocyte is a very thin structure,
less than 1 micrometer thick vs.~a width and a length of about 10-20 micrometers \cite{small}.
The shape of a cell at rest is almost cylindrical, while the shape of a steadily migrating cell
exhibits a simmetry axis. Symmetry arguments and the corresponding high aspect ratios suggest
to represent the cell as a one--dimensional strip, spanning the interval $(x_-(t),x_+(t))$
of the $x$ axis, where the location of the boundaries is to be determined.
In the following, all the relevant physical fields are therefore to be understood as
averaged along the vertical and transverse direction.\\
The onset of polarity in a cell is due to membrane trafficking of a number of proteins.  A
minimal mathematical model able to capture the essential dynamics is provided by two
reaction diffusion equations for a protein (Rho GTPases, for example) that diffuses with different rates
on the membrane and in cytosol \cite{mori}
\begin{align}
\label{0}
\begin{split}
              u_t - D_u u_{xx} &= f(u,w), \\
              w_t - D_w w_{xx} &= -f(u,w),
\end{split}
\end{align}
where $u(x,t)$ and $w(x,t)$ represent the concentration on the membrane and in the cytosol, respectively,
$D_u \ll D_w$, and $f(u,w)$ is typically cubic in $u$. An elementary non--dimensional example is
\begin{equation}
\label{0.1}
      f(u,w) = (u-u_0) (u-u_1) (u-u_2(w+1)).
\end{equation}
Both phases cannot outflow the cell
\begin{equation}
\label{0.2}
               - D_u u_{x} \big|_{x_\pm} =- D_w w_x \big|_{x_\pm} = 0.
\end{equation}
As the cytosolic phase $w$ diffuses much faster than the one attached to the membrane, on the time
scale of interest it has
an homogeneous distribution, slaved to the dynamics of $u$ by conservation arguments.
The system has therefore two stable equilibrium states: a fully homogeneous one, where $u=u_0$,
and a polarized one, when a steady front separates the two regions $u=u_0$ and $u=u_2(\bar w+1)$,
where $\bar w$ depends on the total protein amount and on the physical parameters.
On the basis of this model, the onset of polarization is fully described in terms of active--inactive
protein diffusion and exchange, and the state of polarization of a cell is ``measured'' by the value
of $u$ on its boundary.

Following \cite{larripa}, after vertical and transverse integration the actin cytoskeleton
is here approximated as a one--dimensional viscous fluid with constant contractile
active stress $\alpha$,
\begin{equation}
             \sigma = \mu v_x + \alpha,
\label{1}
\end{equation}
where $\sigma$ is the stress, $\mu$ the shear viscosity, $v(x,t)$ the vertically averaged horizontal velocity
and the subscript denotes differentiation.
While the actual rheology of the cytoskeleton is much more complex \cite{pullarkat, ambrosi},
such a simple model is sufficient to point out some basic mechanisms of cell locomotion.
Notice that both $\sigma$ and $\alpha$ have the dimension of a force,
while $\mu$ is a force multiplied by time.  Moreover, in this simplified framework
the myosin motors are supposed to be uniformly distributed in the cell body, so that $\alpha$ is a constant. \\
The interaction of the cell with the substrate is the shear stress between the cell membrane and the flat
surface, here represented as a drag force proportional to the averaged velocity:
\begin{equation}
                     -  \sigma_x =  - \beta v,
\label{2}
\end{equation}
where $\beta$ is a frictional parameter.
The dynamic boundary conditions in $x_\pm$ account for the tension of the cell membrane \cite{keren2011},
assumed to be proportional to the cell length
\begin{align}
\label{3}
    \sigma_\pm = - \lambda L,
\end{align}
where $L(t) = x_+(t)-x_-(t)$ and we adopt the concise notation $\sigma_+ = \sigma(x_+)$ and so on.
The parameter $\lambda$ is a membrane elastic modulus (force per unit length) \cite{munoz}.\\
While the cell cytoskeleton is made of polymerized actin transported by the velocity $v$,
the monomeric free actin diffuses in the cytosol; the monomers attach to the polymeric filaments
at the cell leading edge, they detach in the body of the cell proportionally to the stress \cite{wilson}, so that
the concentration of G--actin obeys the following reaction--diffusion equation
\begin{align}
                a_t - D a_{xx} = \kappa \sigma,
\label{5}
\end{align}
where $D$ is the monomeric actin diffusivity in cytosol. We assume that near the membrane the
polymerization activity is regulated by the density of G-actin, which is proportional to
the amount of proteins segregated on the membrane $u$:
\begin{align}
\label{5.1}
\begin{split}
                a_\pm = \chi  u_\pm,
\end{split}
\end{align}
where $\chi$ is a constant. According to the dynamics represented by equations \eqref{0} and \eqref{0.1},
at equilibrium two scenarios show up: the boundary conditions are symmetric ($a_+=a_-$) or they do not
($a_+ \neq a_-$), thus accounting for the polarity of the F-actin polymer
\cite{mogilner09}.\\
The polimerized actin is backtransported by the material velocity $v$
\begin{align}
                a^p_t + \left( a^p v \right)_x = - \kappa \sigma.
\label{5.2}
\end{align}
As the cell is a closed system, the total actin is conserved; in other words the
F-actin field is slaved to the G-actin one. Global conservation implies
\begin{align}
    \frac{d}{dt} \int_{x_-}^{x_+} (a+a^p) \, dx =
    \left[ - \dot x (a + a^p) - D a_x + v a^p \right]_{x_-}^{x_+} = 0.
\label{5.3}
\end{align}
The concentration of polimerized actin can be therefore determined a posteriori solving equations
\eqref{5.2} with boundary conditions \eqref{5.3} after that the free actin concentration $a(x,t)$ and
the velocity field $v(x,t)$ have been determined. \\
Assuming that the concentration of the polymeric actin at the edge is fixed, say $a_0$, we get
that the polymerization velocity at the boundary is proportional to the outflux of free
actin,
\begin{align}
\label{6}
\begin{split}
                 - \dot x_+ a_+ - D a_x \big|_{x_+} &=  v^p_+ a_0, \\
                 - \dot x_- a_- - D a_x \big|_{x_-} &=  v^p_- a_0.
\end{split}
\end{align}
The velocity of the cell boundaries $\dot x_\pm(t)$, the material velocity of the cell at the boundary
$v_\pm(t)$ and the actin polymerization rate $v^p_\pm$ (the \textquotedblleft
growth velocity \textquotedblright of the cell) are related by
\begin{align}
\label{4}
       \dot x_\pm = v^p_\pm + v_\pm,
\end{align}
which are the kinematic boundary conditions.
\begin{figure}[h!]
\centering
   \includegraphics[width=14cm]{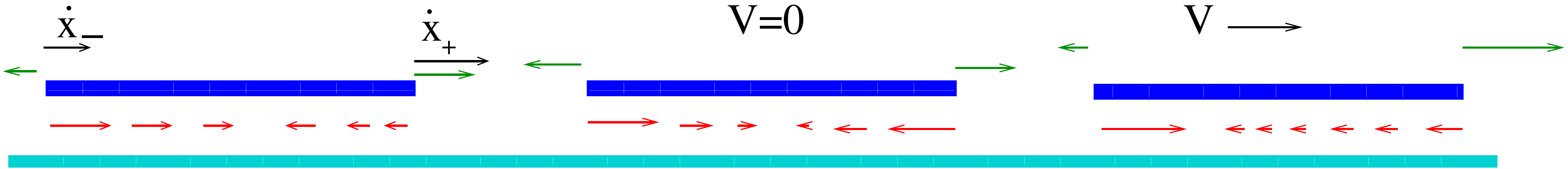}
   \caption{Velocity pattern in a minimal one--dimensional model of a cell (blue) on a flat substrate (light blue).
            The generic non--equilibrium configuration corresponds to a cell length $ L(t) = x_+(t) - x_-(t)$
            moving with speed $V(t) = (\dot x_+(t) + \dot x_-(t))/2$ (left). The two
            possible  equilibrium configurations are the cell at rest (center) and the cell
            migrating with constant speed $V$ and length (right).
            The material and polimerization velocities are plotted in green and red, respectively.
            The cell at rest is characterized by symmetric velocity fields (center),
            while motility emerges from the symmetry breaking (left and right).
            }
\label{fig1d}
\end{figure}

The differential system is now closed: the elliptic-parabolic equations \eqref{2} and \eqref{5}
are supplemented by the 2+2 boundary conditions  \eqref{3} and \eqref{5.1},
whereas the position of the boundaries of the cell are fixed by the kinematic conditions
\eqref{4} supplemented by the constitutive assumption \eqref{6}.

In the next sections we show that the above system of differential equations has
one steady solution and two (symmetric) solutions of travelling wave type,
where the length and the migration velocity of the cell satisfy algebraic
equations that are explicitly stated.

\section{Symmetric solution: cell at rest}
\label{s2}
A steady solution of the differential problem is defined in a constant spatial
domain where all the physical fields are independent of time.
The cell then spans the interval $(-L/2,L/2)$ of the $x$ axis, fixed in time, where $L$ is to
be determined. The solution is expected to be symmetric with respect to the origin, so that
considering the boundary conditions in one boundary point only (say $x=L/2$) is sufficient;
as a matter of fact, $L$ is the only boundary unknown as $\dot x_+=0$.  \\
Under these assumptions equation \eqref{2} can be readily integrated to give
\begin{equation}
               v(x) = - \frac{\alpha+\lambda L}{\sqrt{\beta \mu}}
                        \frac{\sinh\left(\sqrt{\frac{\beta}{\mu}} x \right)}
                             {\cosh\left(\sqrt{\frac{\beta}{\mu}} \frac{L}{2} \right)},
\label{13}
\end{equation}
and
\begin{equation}
       \sigma(x) = \alpha- (\alpha+\lambda L)
                                  \frac{\cosh\left(\sqrt{\frac{\beta}{\mu}} x \right)}
                                       {\cosh \left(\sqrt{\frac{\beta}{\mu}} \frac{L}{2}\right)}.
\label{14}
\end{equation}
\begin{figure}[h!]
\centering
   \includegraphics[width=10cm]{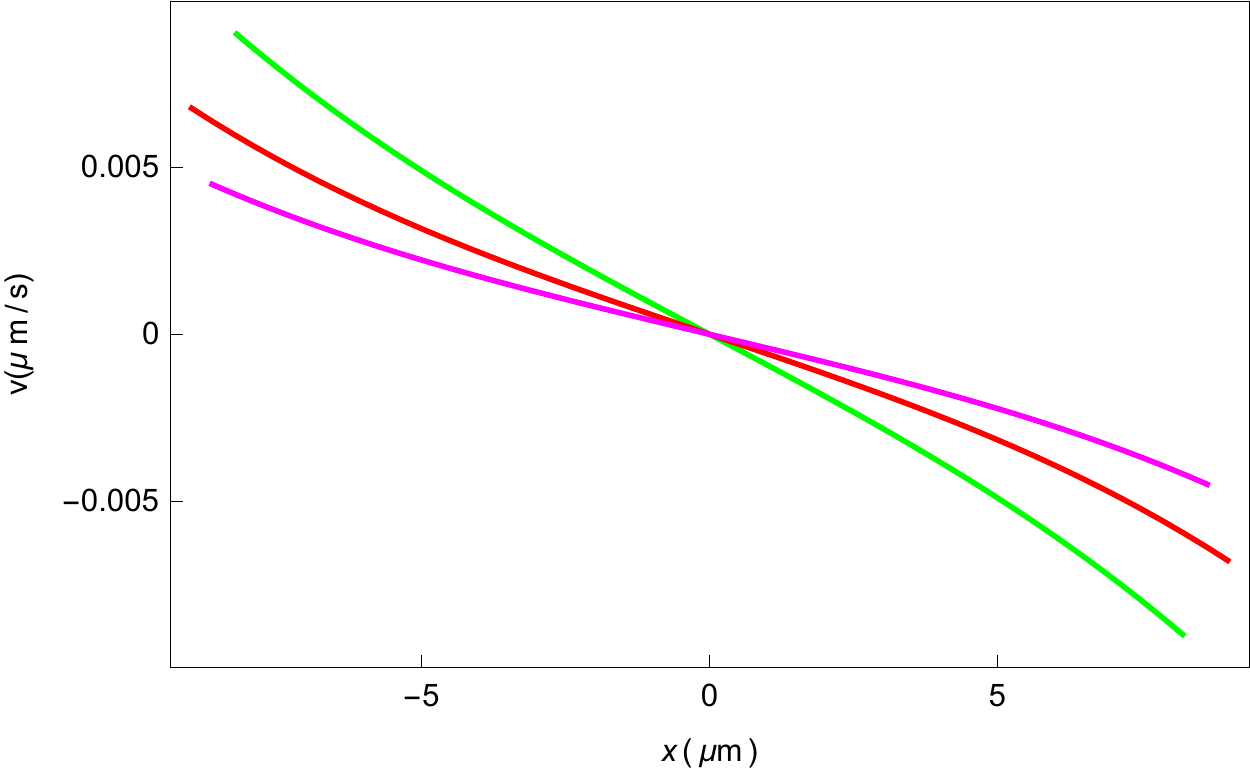}
   \caption{Cell at rest: spatial distributions of F-actin velocity for relevant sets of parameters.
            Green line: $\alpha=0.8\cdot10^3$, $\beta=0.9\cdot 10^4$, $\mu=5.5\cdot10^5$,
             $\lambda=0.5$, $\frac{a_0}{\kappa\mu}=0.33$.
            Red line: $\alpha=0.6\cdot10^3$, $\beta=1.1 \cdot 10^4$, $\mu=5.5\cdot10^5$,
            $\lambda=0.9$, $\frac{a_0}{\kappa\mu}=0.45$.
            Magenta line: $\alpha=0.4\cdot10^3$, $\beta=1.1 \cdot 10^4$, $\mu=5.5\cdot10^5$,
            $\lambda=0.9$, $\frac{a_0}{\kappa\mu}=0.4$.}
\label{figv}
\end{figure}
The plots of F-actin velocity and traction are shown in figures \ref{figv} and \ref{figsigma}
for relevant sets of parameters, over corresponding lengths $L$ (see below). The results
compare well with the experimental data reported for an interval spanning
8 micrometers from the leading edge of the lamellipodium (see figure 2b in \cite{craig}). \\
The actin monomers density can be calculated by integrating equation \eqref{5}, which reduces to
\begin{equation}
                - D a_{xx} = \kappa \sigma.
\label{15}
\end{equation}
As $\sigma(x)$ is an even function and $a_-=a_+$, it follows that $a(x)$ is even, too.
\begin{figure}[h!]
\centering
   \includegraphics[width=10cm]{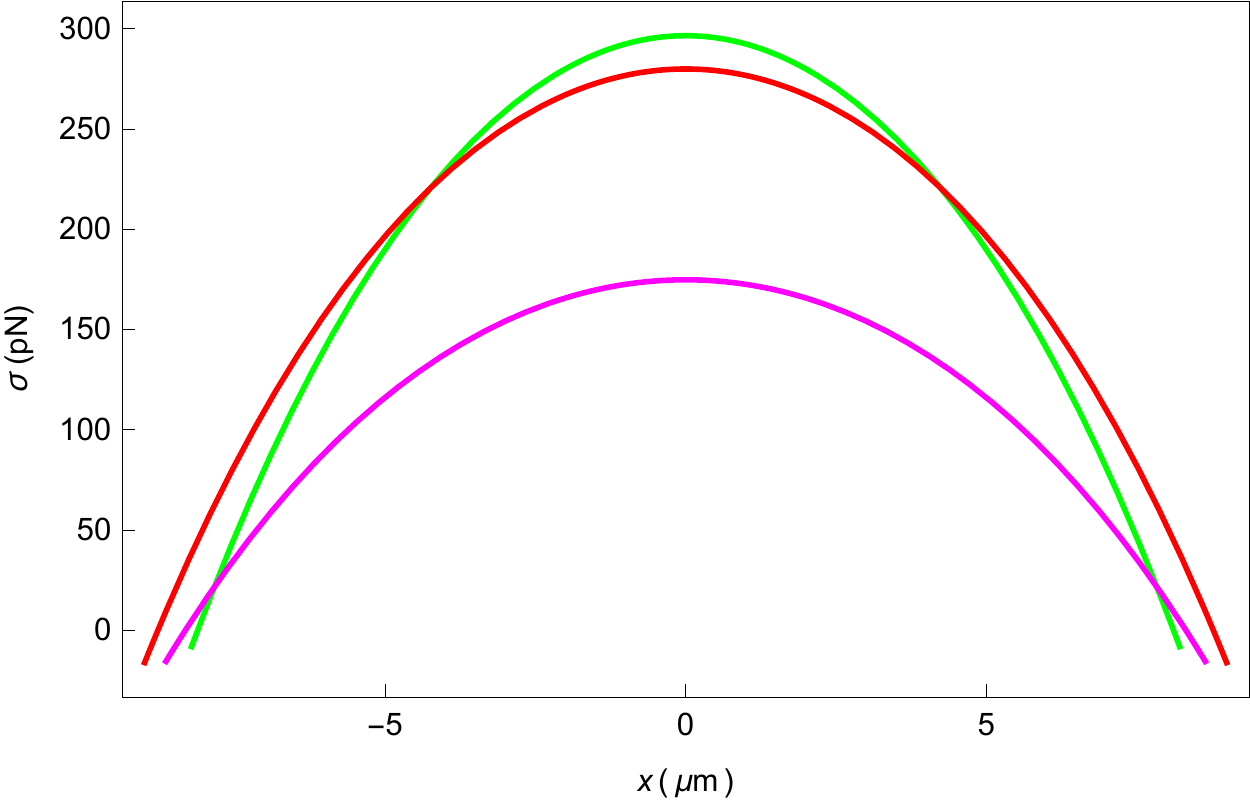}
   \caption{Cell at rest: spatial distribution of traction for relevant sets of parameters.
            Green line: $\alpha=0.8\cdot10^3$, $\beta=0.9\cdot 10^4$, $\mu=5.5\cdot10^5$,
            $\lambda=0.5$, $\frac{a_0}{\kappa\mu}=0.33$.
            Red line: $\alpha=0.6\cdot10^3$, $\beta=1.1 \cdot 10^4$, $\mu=5.5\cdot10^5$,
            $\lambda=0.9$, $\frac{a_0}{\kappa\mu}=0.45$.
            Magenta line: $\alpha=0.4\cdot10^3$, $\beta=1.1 \cdot 10^4$, $\mu=5.5\cdot10^5$,
            $\lambda=0.9$, $\frac{a_0}{\kappa\mu}=0.4$.}
\label{figsigma}
\end{figure}

To determine the actin concentration at the boundary we integrate once equation \eqref{15}
\begin{equation}
                - D a_{x} = \kappa g(x),
\label{18}
\end{equation}
where $g(x):=\int_{0}^{x} \sigma$.
Using the boundary condition
\eqref{6} and the constitutive equation \eqref{6} we get
\begin{equation}
                 \kappa g_+ = a_0 v^p_+.
\label{19}
\end{equation}
Using the known form of the stress \eqref{14}, equation \eqref{19} explicitly rewrites as follows
\begin{equation}
                  a_0 v^p_+
                 = \kappa \left(
                                 \alpha \frac{L}{2}- (\alpha+\lambda L) \sqrt{\frac{\mu}{\beta}}
                                  \tanh\left(\sqrt{\frac{\beta}{\mu}} \frac{L}{2} \right)
                          \right).
\label{19.1}
\end{equation}
The kinematic condition \eqref{4} then provides an algebraic equation for the length of the
cell at rest
\begin{equation}
                           v_+ + \frac{\kappa}{a_0} g_+ = 0,
\label{19.2}
\end{equation}
or explicitly,
\begin{equation}
      \alpha \frac{L}{2}
       - (\lambda L +\alpha) \sqrt{\frac{\mu}{\beta}} \left( 1 + \frac{a_0}{\kappa \mu} \right)
         \tanh \left(\sqrt{\frac{\beta}{\mu}}\frac{L}{2}\right) =0.
\label{20}
\end{equation}
The physical meaning of equation \eqref{19.2} is the following: given $v$ and
$g$ as functions of the cell length, the boundary of the cell is fixed at $L/2$ such that the
retrograde flow exactly balances the outflow of free actin
(which rereads as elongation rate of the network polymerized actin).
The same argument has been enforced by Etienne et al.: a competition of retrograde flow
and protrusion, modulated by the stiffness of the environment, is the means by which cell
area is regulated \cite{etienne}.
Because of the symmetry of the problem, the result \eqref{19} is notably independent
of the boundary conditions \eqref{6} that are instead expected to play a role when the cell migrates.

There exists a unique, positive solution $L$ of equation \eqref{20} if $\lambda$
is sufficiently small, i.e. if the parameters satisfy the following inequality (see Appendix A)
\begin{equation}
\lambda <\frac{\alpha \kappa}{2}\frac{\sqrt{\beta \mu}}{(a_0+\kappa \mu)}.
\label{lambda}
\end{equation}
The existence condition \eqref{lambda} can be rewritten in a more perspicuous way
\begin{equation}
          \left(1+\frac{a_0}{\kappa \mu} \right) \lambda \frac{\mu}{\beta} < \frac{\alpha}{2}.
\label{lambda1}
\end{equation}
The quantity $a_0/\kappa \mu$ represents the ratio between the material velocity $v$
and the polymerization velocity $v_p$ and it is known to be smaller than one. Therefore,
neglecting the term in the bracket at the left hand side, condition \eqref{lambda1} requires
that the membrane tension, evaluated using the representative viscous decay length $\sqrt{\mu/\beta}$
is much smaller than the active stress produced by the actomiosin machinery, a regime
that definitely applies in a fish keratocyte.

\begin{figure}
        \centering
        \begin{subfigure}{0.55\textwidth}
                \includegraphics[width=\textwidth]{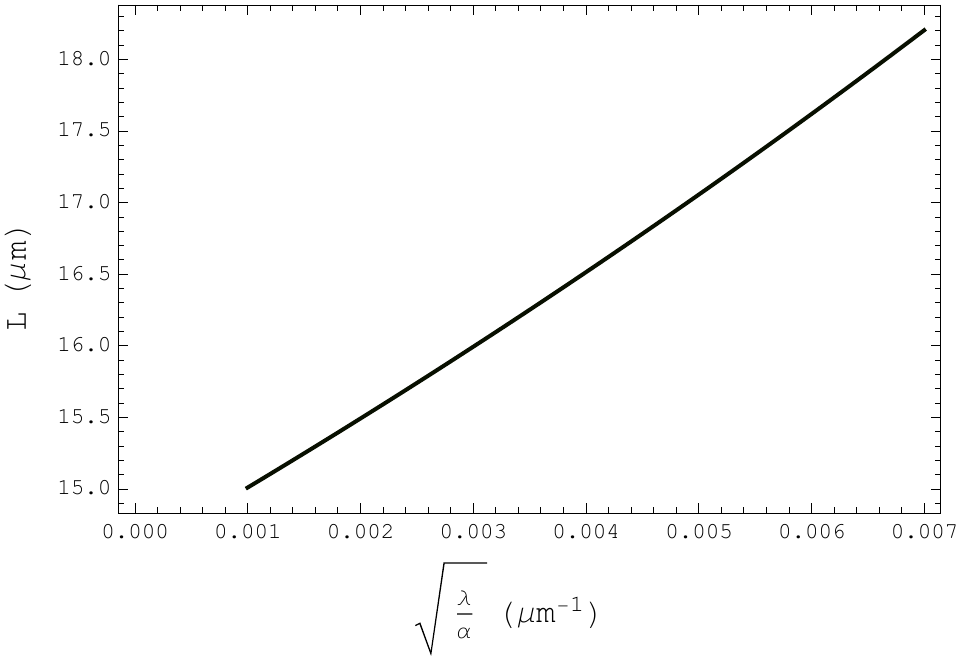}
                \caption{}
                \label{fig:lambda}
        \end{subfigure}%
        \begin{subfigure}{0.55\textwidth}
                \includegraphics[width=\textwidth]{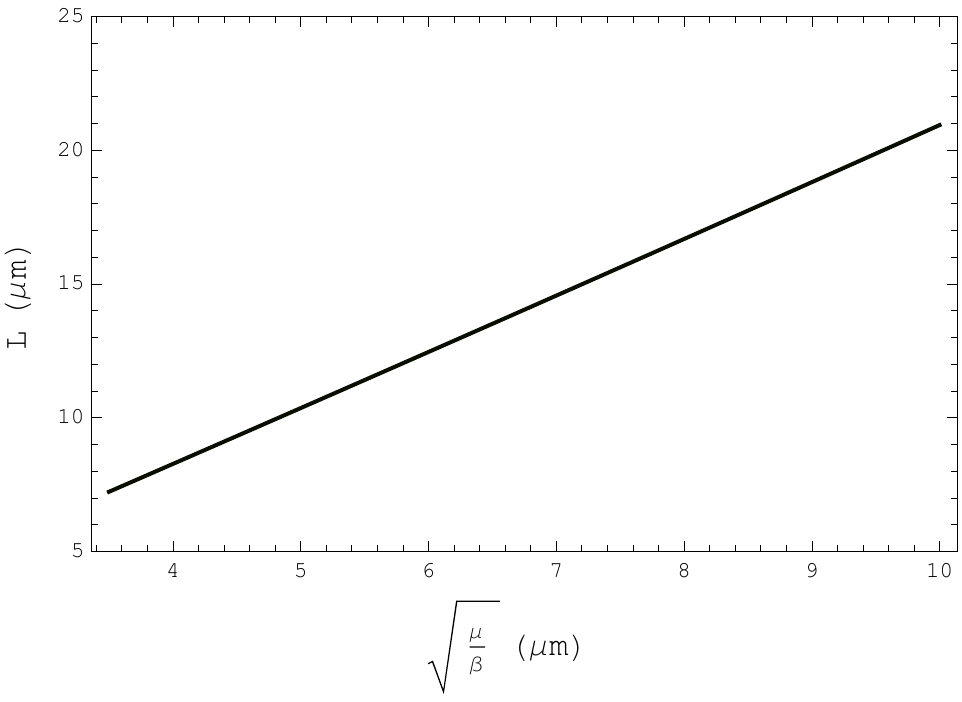}
                \caption{}
                \label{fig:mu}
        \end{subfigure}

        \centering
        \begin{subfigure}{0.55\textwidth}
                \includegraphics[width=\textwidth]{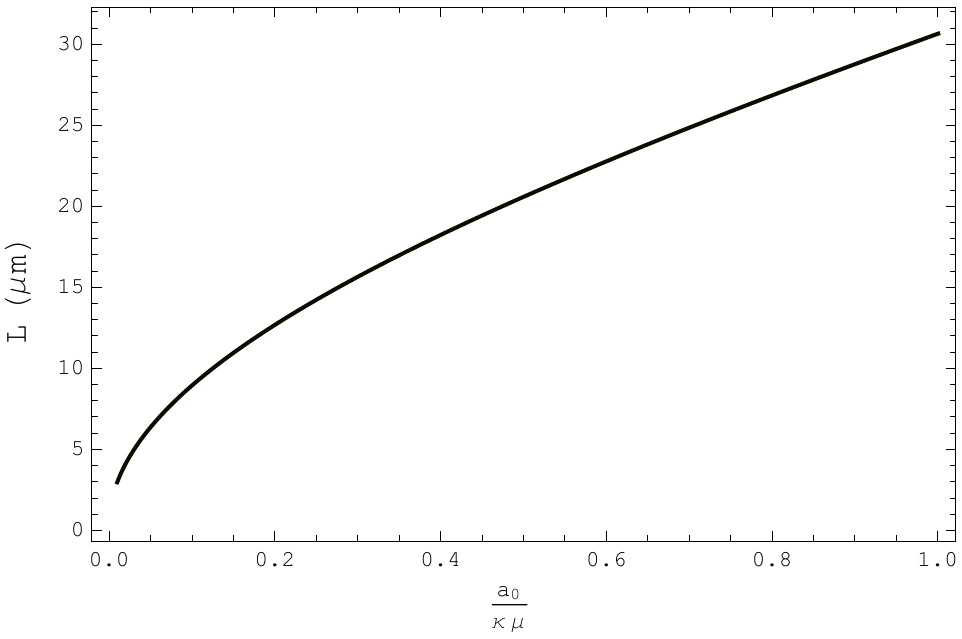}
                \caption{}
                \label{fig:k}
        \end{subfigure}
        \caption{Plots of the cell length as a function of the relevant parametric
                 combinations
                 (a) $\frac{\lambda}{\alpha}$, (b) $\sqrt{\frac{\mu}{\beta}}$,
                 (c) $\frac{a_0}{\kappa \mu}$.
                 The fixed parameters take the values given
                 in table~\ref{tab1}. In  Figure~\ref{fig:parameters}
                 (a) and Figure~\ref{fig:parameters}(b) $\frac{a_0}{\kappa \mu}=0.33$.}
                \label{fig:parameters}
\end{figure}

A validation of the above theory comes from a comparison between the observed cell length and the one
predicted according to \eqref{20}, where physical measured parameters are employed.
The values of the most relevant parameters used in the model are listed in Table~\ref{tab1}, as
extracted from the literature.
On the basis of the smallness argument $a_0/\kappa \mu < 1$ illustrated above,
the values of $a_0$ and $\kappa$ have minor relevance.
\begin{table}[h!]
    \begin{center}
    \begin{tabular}{| l | l | l | l |}\hline
    Parameter & Physical Meaning & Value & Source \\\hline
    $\alpha$  & contractile active stress & $10^3 \,\rm{pN}$                                 & \cite{larripa}\\\hline
    $\beta$   & friction coefficient      & $10^4 \, \rm{(pN\, s)/\mu m^2 }$                 & \cite{larripa}\\\hline
    $\mu$     & viscosity coefficient     & $5\cdot 10^5 \,\rm{pN\,s}$                       & \cite{kruse} \\\hline
    $\lambda$ & membrane elastic modulus  & $\lambda=0.5 \,\rm{pN/\mu m} $                   & \cite{munoz} \\\hline
    $D$       & diffusion coefficient     &  $  10 \,\rm{\mu m^2/s}$                         &\cite{novak}  \\\hline
    \end{tabular}
    \caption{Model parameters based on literature referred to fish keratocytes.}\label{tab1}
    \end{center}
\end{table}
Figure~\ref{fig:parameters} describes how the cell length changes when the model parameters are varied.

By using the parameters value extracted from the literature and reported in Table~\ref{tab1}
we numerically compute $L\backsimeq 18 \mu m$ which is consistent with the experimental results
on fish keratocytes \cite{vallotton}.

\begin{figure}[h!]
\centering
\includegraphics[width=9cm]{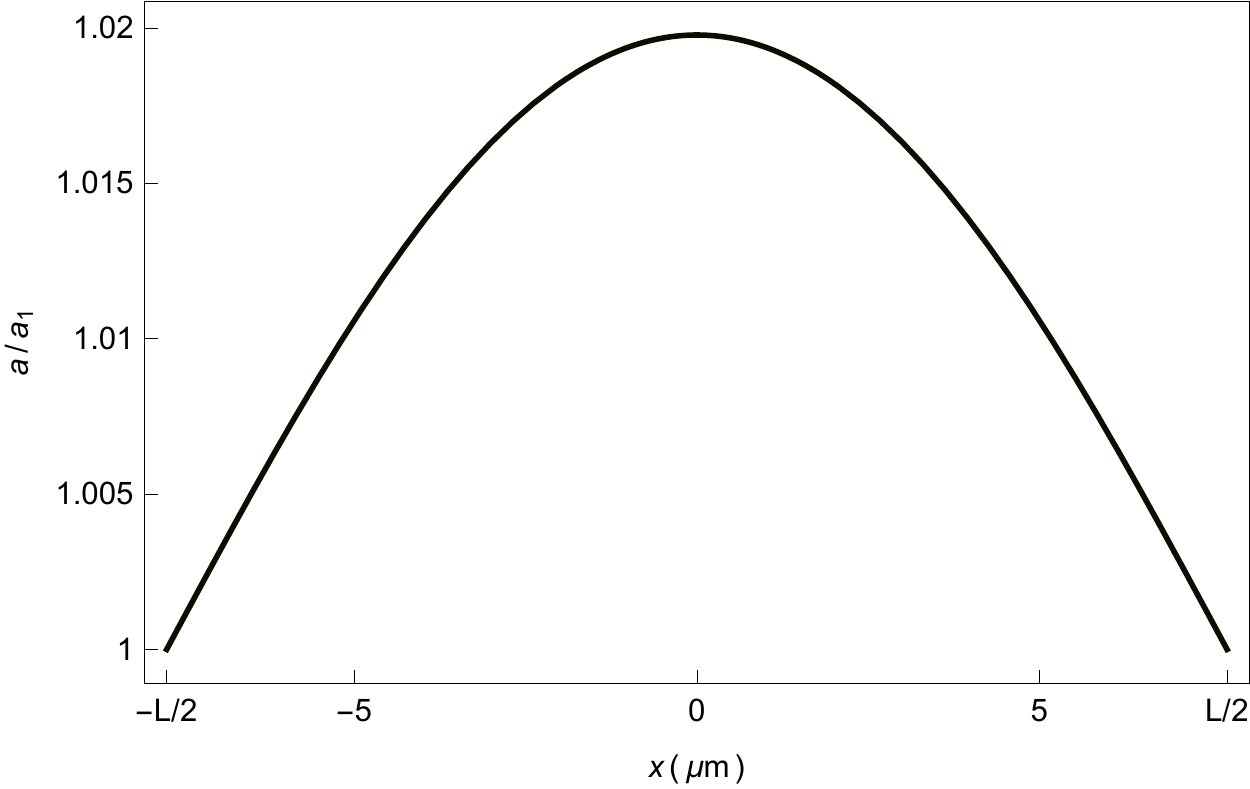}
\caption{Cell at rest: spatial distribution of monomeric actin
         for values of model parameters listed in table~\ref{tab1}.
         Here $a_0/a_1=3.6$; $a_0/\kappa \mu=0.36$.
         }
\label{fig8}
\end{figure}

The G-actin density field can be obtained solving equation \eqref{15} with symmetric boundary conditions \eqref{5.1}
(say $a_\pm = \chi u_1 = a_1$) thus giving
\begin{align}
a(x)=& \frac{\kappa}{D} \frac{\mu}{\beta} (\lambda L +\alpha)
       \left( \frac{\cosh\left(\sqrt{\frac{\beta}{\mu}} x \right)}
                   {\cosh\left(\sqrt{\frac{\beta}{\mu}} \frac{L}{2}\right)}
              -1 \right)
    + \frac{\kappa}{D} \frac{\alpha}{2} \left(\frac{L^2}{4}-x^2\right)+ a_1.
\end{align}
Figure~\ref{fig8} shows the free actin profile for a cell at rest. It is symmetric
and the free actin concentrates at the center of the cell in correspondence
to the maximum stress concentration, in agreement with observations.

\section{Non-symmetric solution: polarized migrating cell}
\label{s3}
In this section we look for travelling wave type solutions of the system illustrated in Section \ref{s1}.
Our conjecture is that the polarity of the cell can be represented by an asymmetry in the boundary conditions
for G-actin: the imbalance in actin polimerization at the boundaries drives the motion. To fix
the ideas, we assume that the steady state solution of equation \eqref{5.1} is $a_+=a_1, a_-=a_2$.
All the fields are here supposed to depend on $x$ and $t$ in terms of the combination
$y=x-Vt$ where $V$ is a constant to be determined; if a solution $V\neq 0$ exists, it corresponds to a motile cell.
Then we assume
$$
     y_-=x_- -Vt=-L/2, \qquad y_+=x_+-Vt=L/2,
$$
with $L,V$ to be determined.
The force balance equation involves no derivatives in time and the stress depends only on the
spatial variations of the velocity, so that all the calculations of the previous sections
and the results \eqref{13} and \eqref{14} for the velocity and stress field remain valid.
We notice that $v(x)$ given by equation \eqref{13} is the velocity field of the cytoskeleton measured
by an observer travelling with constant velocity $V$, while an observer at rest records $v(x)+V$. \\
The density of monomeric actin can now be calculated by solving equation \eqref{5}, which here rewrites
\begin{equation}
              -V a'  - D a'' = \kappa \sigma.
\label{13.9}
\end{equation}
In order to determine the actin concentration we integrate once equation \eqref{13.9}
\begin{equation}
              -Va  - D a' = \kappa g + c_1,
\label{13.12}
\end{equation}
for which no symmetry argument in principle applies. \\
We look for solutions of equation \eqref{13.12} of the type
$$
               a(y) = h(y) \e^{-\frac{V}{D} y},
$$
and we find
\begin{equation}
              a(y) = c_2 \e^{-\frac{V}{D} y} -\frac{c_1}{V}
                     - \frac{\kappa}{D} \e^{-\frac{V}{D} y}
                    \left( \int \e^{\frac{V}{D} y} g \right),
\label{13.13}
\end{equation}
where $c_1,c_2$ are constants to be fixed enforcing the boundary conditions \eqref{5.1}.
In particular, we get
\begin{align}
\label{20.1}
\begin{split}
c_1 = - \frac{V}{2\sinh\left({\frac{V}{D} \frac{L}{2}}\right)}
       &  \left\{ \frac{\kappa}{D}
               \left[ \left( \int \e^{\frac{V}{D} y} g \right)_{L/2}
                     -\left( \int \e^{\frac{V}{D} y} g \right)_{-L/2}
               \right] \right. \\
         &   \left. + a_1 \e^{\frac{V}{D} \frac{L}{2}}
              - a_2 \e^{-\frac{V}{D} \frac{L}{2}}
        \right\}
\end{split}
\end{align}
where $V$ and $L$ are to be determined on the basis of the kinematic boundary conditions.

\begin{figure}[h!]
\centering
\includegraphics[width=12cm]{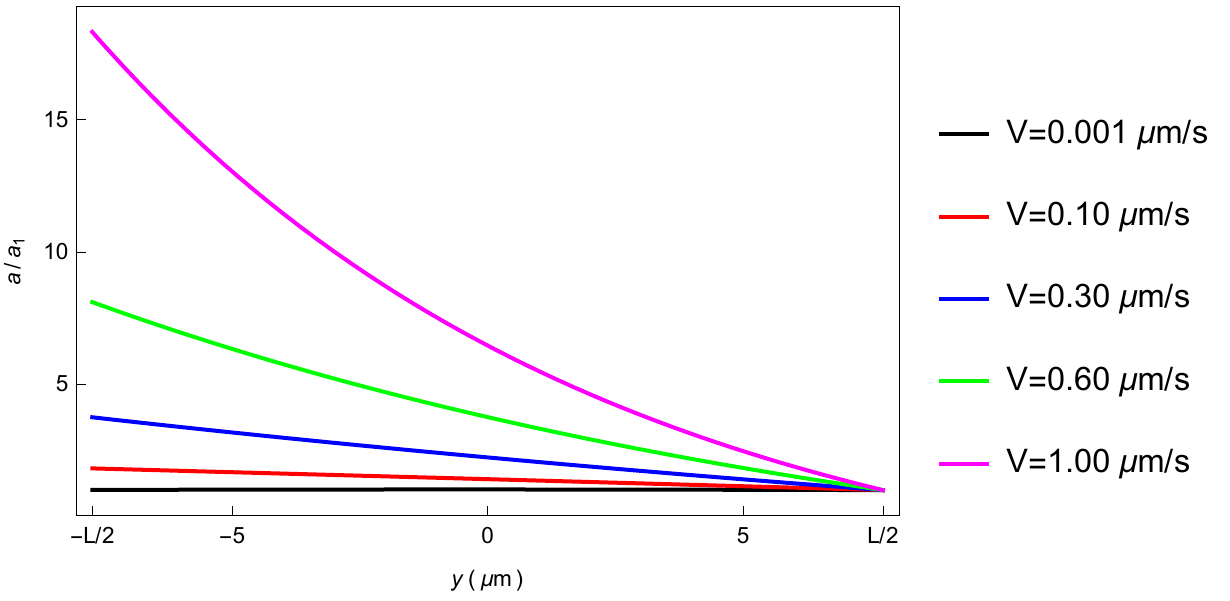}
\caption{Migrating cell: spatial distribution of monomeric actin for a cell moving
         in the positive $x$ direction with different (given) migration velocities.
         Actin distributions are normalized using the value at the front edge:
         $a(L/2)=a_1=5\cdot10^6$, $a(-L/2)=a_2$.
         We set $a_2=1.0078\,a_1$ (black line), $a_2=1.82\,a_1$ (red line), $a_2=3.75\,a_1$
         (blue line), $a_2=8.1\,a_1$ (green line), $a_2=18.3\,a_1$ (magenta line).
         Moreover $a_0/\kappa \mu=0.36$.
         The remaining model parameters take the values listed in table~\ref{tab1}.}
\label{fig11}
\end{figure}
\begin{figure}[h!]
\centering
\includegraphics[width=12cm]{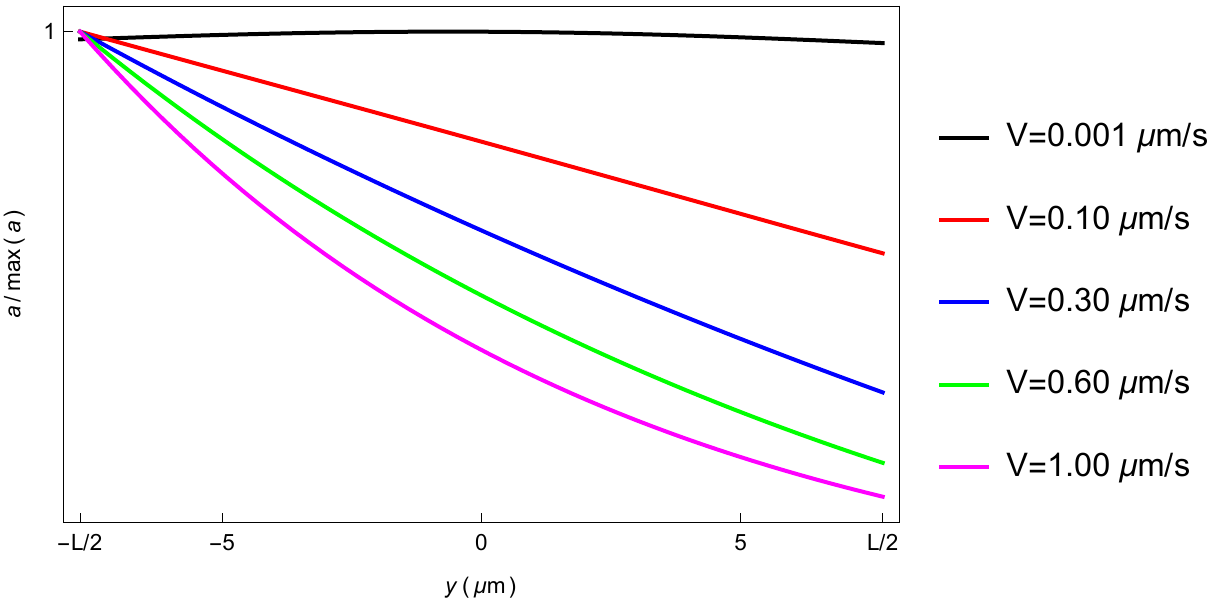}
\caption{Migrating cell: spatial distribution of monomeric actin for a cell moving
         in the positive $x$ direction with different migration velocities.
         Actin distributions are normalized by their maximal absolute values.
         In each computation $a(L/2)=a_1$, $a(-L/2)=a_2$ with with $a_1=5\cdot10^6$.
         We set $a_2=1.0078\,a_1$ (black line), $a_2=1.82\,a_1$ (red line),
         $a_2=3.75\,a_1$ (blue line), $a_2=8.1\,a_1$ (green line), $a_2=18.3\,a_1$ (magenta line).
         Moreover $a_0/\kappa \mu = 0.36$. The remaining model parameters take
         the values listed in table~\ref{tab1}.}
\label{fig12}
\end{figure}

The kinematic boundary conditions in a travelling frame of reference follow \eqref{6} and \eqref{4}
when $\dot x_+ = V$
\begin{align}
\label{22}
\begin{split}
                 - V a_+ - D a_x \big|_{L/2}  &=  (V-v_+) a_0, \\
                 - V a_- - D a_x \big|_{-L/2} &=  (V-v_-) a_0.
\end{split}
\end{align}
Since $v(y)$ is an odd function, subtracting the latter equations
we find that the cell length is provided by
\begin{equation}
                2 v_+ + \kappa \int_{-L/2}^{L/2} \!\!\!\! \sigma \,\,dy =0
\label{24.1}
\end{equation}
and we again recover the condition \eqref{19.2}, which is independent of $V$. \\
In order to determine the velocity of the cell we sum equations \eqref{22} and get
\begin{align}
             \left(- V a - D a_x \right)_{L/2} + \left(- V a - D a_x\right)_{-L/2}
                               = 2 V a_0. \label{25}
\end{align}
The physical meaning of equation \eqref{25} is transparent: motion can be produced
only by an imbalance of actin flow at the border. We recall the form of the fluxes \eqref{6}:
by using the integral \eqref{13.12} and the fact that $g$ is an odd function we get
\begin{align}
    a_0 V = c_1.
\end{align}
or, using equation \eqref{20.1}
\begin{align}
a_0 V =  \frac{-V}{2\sinh\left({\frac{V}{D} \frac{L}{2}}\right)}
         \left\{ \frac{1}{D}
               \left[ \left( \int \e^{\frac{V}{D} y} g \right)_{L/2}
                   \!\!   - \left( \int \e^{\frac{V}{D} y} g \right)_{-L/2}
               \right]
             + a_2 \e^{\frac{V}{D} \frac{L}{2}}
              - a_1 \e^{-\frac{V}{D} \frac{L}{2}}
        \right\}
\end{align}
After discarding the $V=0$ solution, corresponding to the case $c_1=0$ explored in the previous section,
we finally determine the condition that fixes the velocity of the travelling cell
\begin{align}
\label{26}
  2 a_0 \sinh\left({\frac{V}{D} \frac{L}{2}}\right) = &
         \frac{- 1}{D}
               \left[ \left( \int \e^{\frac{V}{D} y} g \right)_{L/2}
                     \!\!-  \left( \int \e^{\frac{V}{D} y} g \right)_{-L/2}
               \right]
             - \left( a_2 \e^{\frac{V}{D} \frac{L}{2}} - a_1 \e^{-\frac{V}{D} \frac{L}{2}} \right)
\end{align}
where $L$ is the unique solution of equation \eqref{24.1}. In Appendix B it is shown
that for $a_2>a_1$ the algebraic equation \eqref{26} has a unique positive solution $V$;
viceversa $-V$ is solution when $a_1$ and $a_2$ are exchanged.
\begin{figure}[h!]
\centering{
        \begin{subfigure}{0.45\textwidth}
                \includegraphics[width=\textwidth]{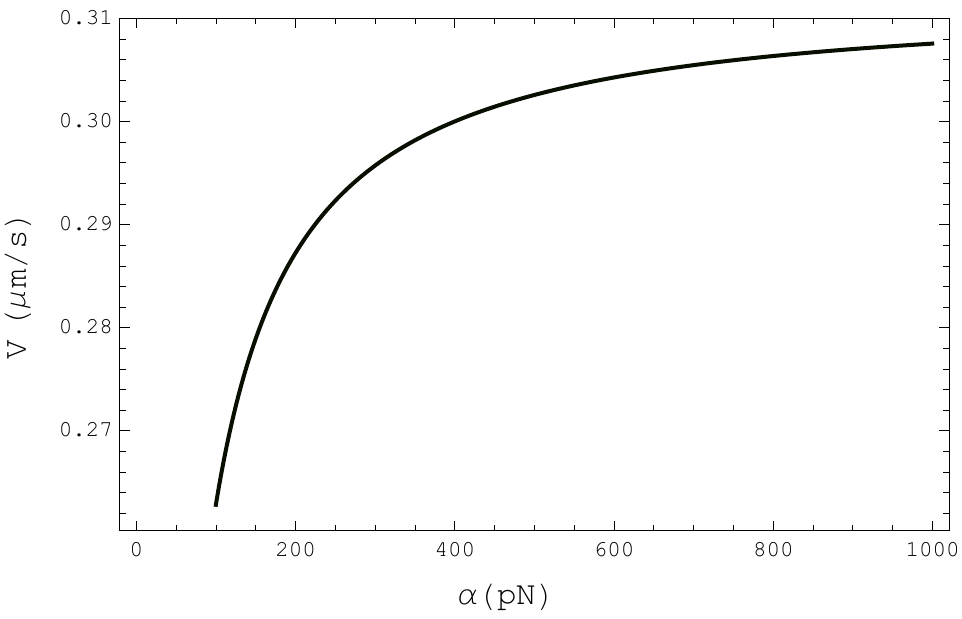}
        \caption{}
        \end{subfigure}
       \begin{subfigure}{0.43\textwidth}
                \includegraphics[width=\textwidth]{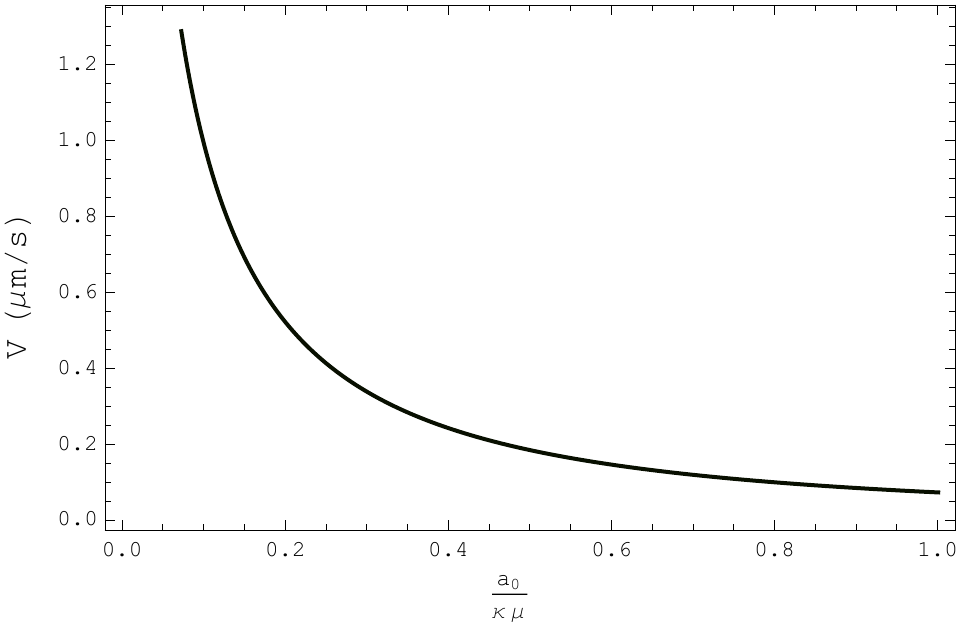}
        \caption{}
        \end{subfigure}
        \begin{subfigure}{0.45\textwidth}
                \includegraphics[width=\textwidth]{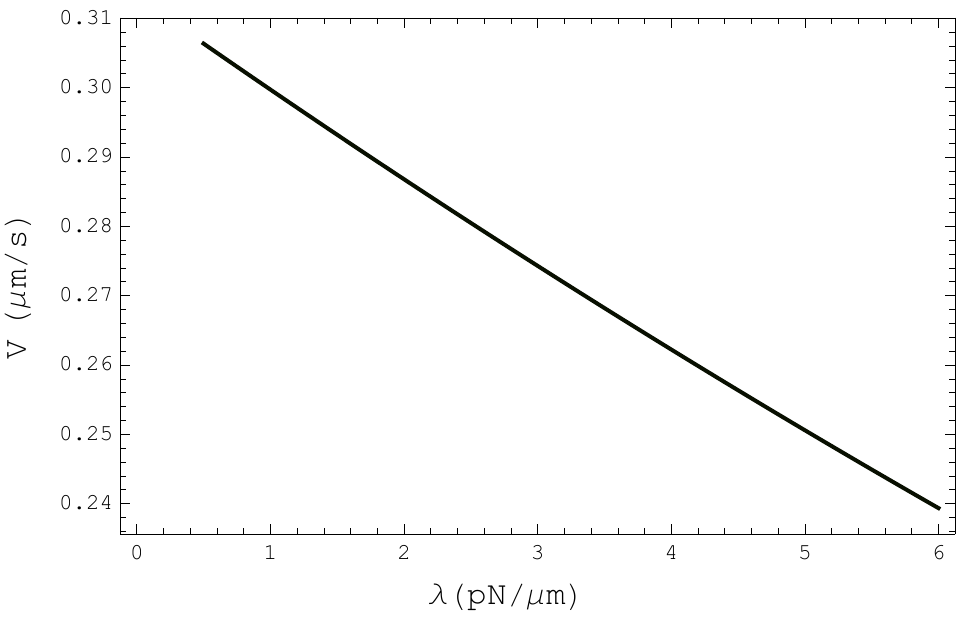}
       \caption{}
       \end{subfigure}
       \begin{subfigure}{0.44\textwidth}
                \includegraphics[width=\textwidth]{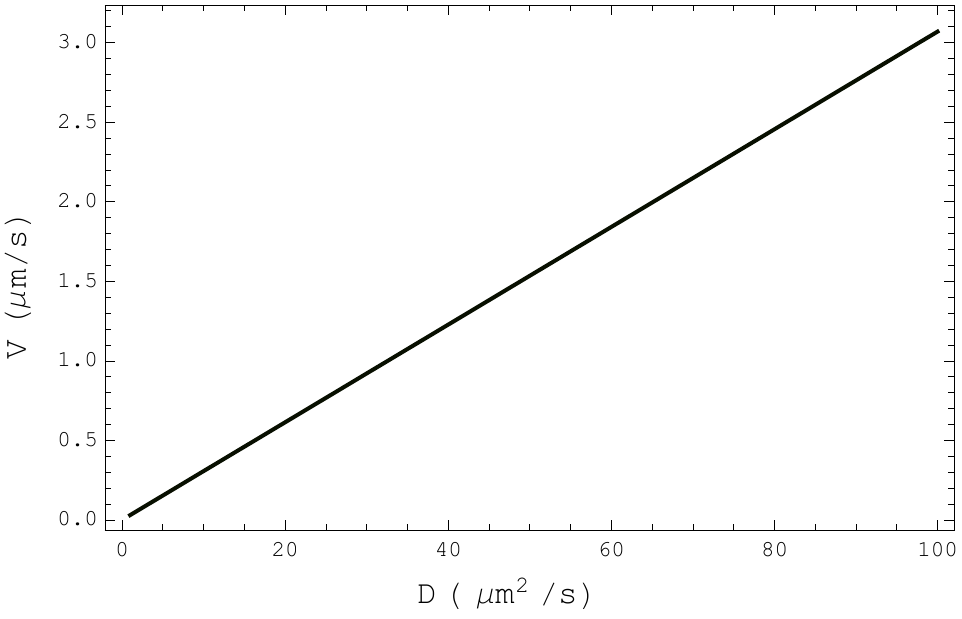}
        \caption{}
        \end{subfigure}
        \begin{subfigure}{0.45\textwidth}
                \includegraphics[width=\textwidth]{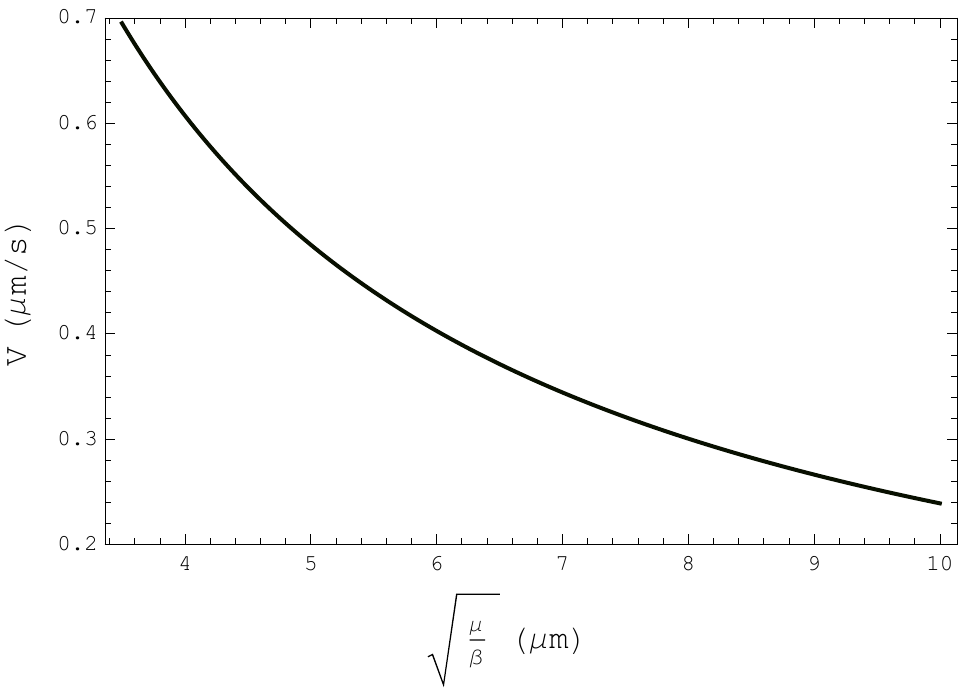}
\caption{}
        \end{subfigure}}
        \caption{Cell velocity as a function of: (a) $\alpha$,
                 (b) $\frac{a_0}{\kappa \,\mu}$  , (c) $\lambda$, (d) $D$, (e) $\sqrt{\frac{\mu}{\beta}}$.
                In each computation the fixed parameters take the values
                given in table~\ref{tab1}. Moreover, we set
                $a(\frac{L}{2})=a_1=5\cdot10^6$, $a(-\frac{L}{2})=a_2=4a_1$
                and in (a), (b), (c), (d) $\frac{a_0}{\kappa \mu}=0.33$.}
\label{fig:parameters_V}
\end{figure}

In Figure \ref{fig:parameters_V} the migration velocity predicted by equations
(\ref{24.1}) and (\ref{26})  is plotted versus
some relevant parametric combinations. The migration velocity ranges the interval
0.1-0.3 micrometers per second, consistently with the observations
\cite{zimmermann2012, vallotton, jurado}.
As could be expected, $V$ diminishes linearly with
$\lambda$ and grows linearly with the G-actin diffusion coefficient $D$, but it is damped
more than linearly versus the viscous diffusion length $\sqrt{\mu/\beta}$. The dependence
of $V$ versus the active stress and versus the ratio between material velocity and
growth velocity provide remarkable information, that could not be foreseen without
the support of a mathematical model.
In fact, $V$ grows versus $\alpha$ according to a curve that saturates
at $V\simeq .3 {\mathrm \,\,\mu m \,\, s^{-1}}$ around the value of 400 pN,
thus suggesting that the migration velocity weakly depends on the active stress.
Conversely, the very rapid growth of $V$ when $a_0/\kappa \mu \rightarrow 0$, prompts the
conjecture that the retrograde velocity plays a stabilizing role in cell mechanics,
keeping the value of $V$ in the portion of the curve at the right of the parametric value
$a_0/\kappa \mu \simeq 0.2$.

According to equation \eqref{24.1}, a migrating fish keratocyte has the same size of a
cell at rest; while this is physically nearly true, we observe that such an approximation
is due to the simplicity of our model, that does not account for
the stress field generated by an asymmetric pattern of myosin. Our approach is therefore
complementary to Recho et al \cite{recho15}, who focus on the initiation of motility
as driven by contraction only. While the introduction
of a dependence of the stress field on the myosin concentration in equation \eqref{1}
would be straightforward, the solution of the equations should be written in terms
of special functions, and the final results would not be that transparent. More important,
equation \eqref{26} suggests that the key ingredient to sustain the cell motility
is the imbalance of actin flow between leading and trailing edge.
The (symmetric) stress field however plays a fundamental role in determining the actin
depolimerization rate, which is maximum far from the lamellipodium edge.
While actin, myosin and adhesion orchestrate the transition from (apparently) static
to migrating cell \cite{barnhart}, the nonlinearity of the actin flux at the boundaries,
that reflects the polarity of the polymeric chain, is sufficient to feed
the locomotion machinery of the polarized cell.

\begin{figure}[h!]
\centering
\includegraphics[width=10cm]{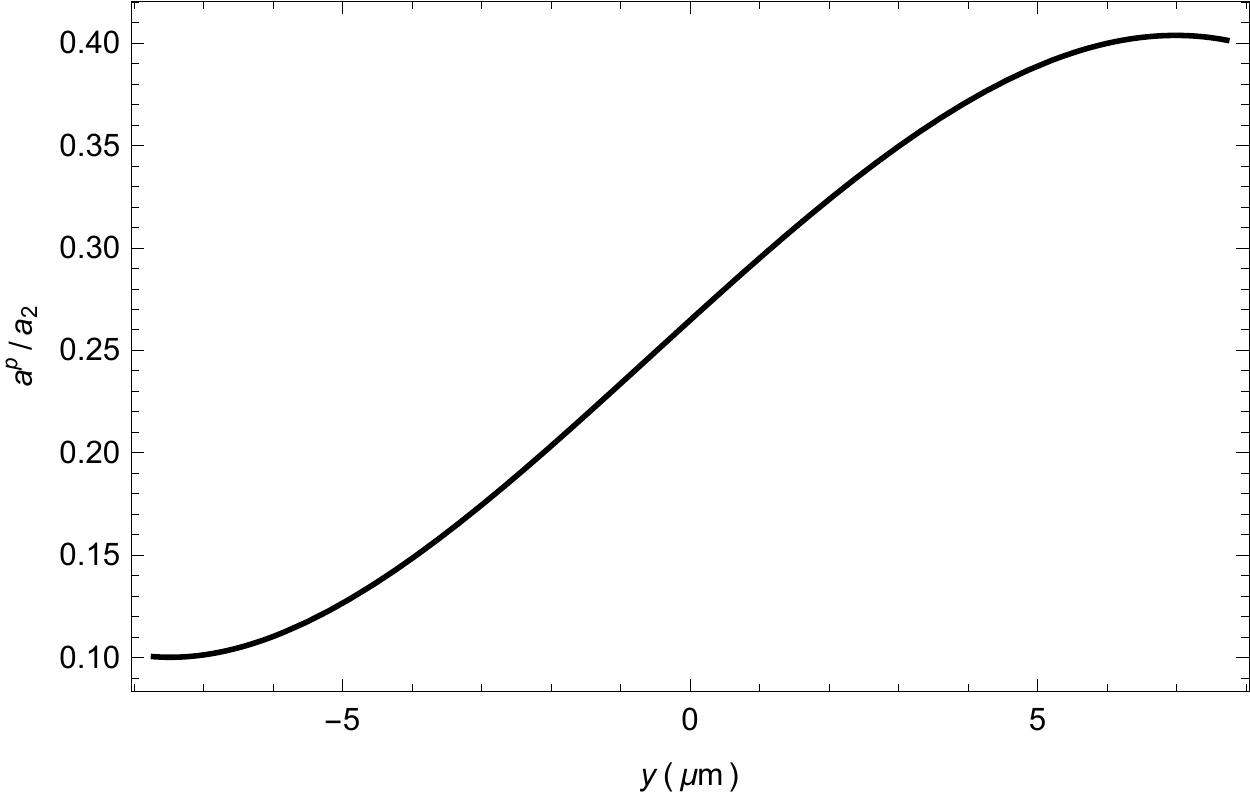}
\caption{Migrating cell: spatial distribution of polimeric actin for a cell moving
         in the positive $x$ direction according to \eqref{29}.
         Actin distributions is normalized by its maximal value. All the
         parameters are as listed in table~\ref{tab1}, corresponding to L=15.4 micrometers
         and V=0.11 micrometers per second.
         }
\label{fig_ap}
\end{figure}

As observed in Section \ref{s1}, the profile of F-actin $a^p(y)$ can now be calculated
a posteriori.  In a travelling frame of reference equation \eqref{5.2} rewrites
\begin{align}
                \left( (v-V) a^p \right)_y = - \kappa \sigma.
\label{27}
\end{align}
As we interested in the physical regime $|V|>|v|$, the quantity $v-V$ always has the sign of $-V$ on the
boundary. Assuming, for instance, that $V>0$, the characteristics enter the right boundary and
therefore equation \eqref{27} has to be integrated in $[y,L/2]$,
\begin{align}
                (v-V) a^p = \left[(v-V) a^p\right]_{L/2} + \kappa \int_y^{L/2} \sigma
\label{28}
\end{align}
or,
\begin{align}
                a^p(y) = \frac{-\left[-V a - D a' \right]_{L/2} + \kappa \int_y^{L/2} \sigma}{v-V},
\label{29}
\end{align}
where the boundary condition (\ref{22}a) has been used. The profile of F-actin according to equation
\eqref{29} is plotted in figure \ref{fig_ap}.

\section*{Final remarks}
A fragment of lamellipodium of fish keratocyte is a very attractive minimal biophysical model
to investigate the physics of cell motility.  We have formulated a mathematical model
that accounts for the basic mechanisms that drive the crawling of a lamellipodial fragment.
While polarization of the cell is governed by a bistable reaction--diffusion equation
accounting for the trafficking of proteins on the plasma membrane and in the cytosol,
the concentration of active proteins dictates the polimerization
rate at the edges of the cell. Our contribution is in coupling this reaction--diffusion
system with mechanics: the stress modulates the F-actin depolimerization rate in bulk
while the exchange between actin phases near the membrane is governed by the concentration
of active proteins. In particular, in a polarized cell, the imbalance in the exchange
between F-actin and G-actin produces the displacement.

The mathematical model illustrated in this work has a number of physical
and biological limitations.  Traction force microscopy reveals that when a fish keratocyte
travels, the largest stress occurs in the transverse direction \cite{fournier};
this compressive pattern, which is probably instrumental to the stabilization of
the trajectory and to the rearward depolimerization of the cytoskeletal network, is ruled out by
our transverse average. Moreover the active stress is here supposed to be independent of
the position, while information about the non-homogeneous distribution of myosin motors are known.
Last, but not least, the polarization dynamics and the stress pattern are here fully decoupled;
this is probably true in the specific regimes we consider, but the reorganization
of the actin network observed when the cell passes from the state at rest to the motile one clearly
indicates that forces that play a role. The same argument applies when a cell has to produce a force
against an obstacle to move, a framework not considered here but which is clearly of paramount
relevance in the perspective of 3D migration. In the particular case of a stalled cell,
the retrograde velocity is approximately equal to the polimerization velocity
of a free cell \cite{zimmermann2012}. This demonstrates that in more complex scenarios,
when external loads are applied, the mechanics produces a feedback on the polimerization rate
that is not captured by the present model.

Notwithstanding only few physical ingredients have been introduced, a number of analytical results
have been obtained.  The linear equations representing the balance of mass of free actin
and the balance of momentum for the polymerized actin, can be integrated, thus providing profiles
of concentration, velocity and stress that compare well with the observed ones.
All the nonlinearity of the system, needed to account for the observed bistability, is encoded
in the bistable reaction--diffusion system that govern the phase exchanges
(oligomeric vs.~polymeric) at the boundary,
i.e. in the kinematic and dynamic boundary conditions. The use of linear constitutive equations
allows to find a closed (algebraic) form for the dependence of the length and speed of a fragment
on the parameters of the model which account for the observed bistability. Using
values of the physical parameters taken from the literature, we find that our predictions
agree with the reported ranges, thus backward supporting the validity of the overall theoretical model.

\subsection*{Acknowlegments}
We are indebted with Paolo Biscari and Pasquale Ciarletta for fruitful discussions
about the content of this paper. This work was partially supported by the
“Start-up Packages and PhD Program” project, co-funded by Regione Lombardia
through the “Fondo per lo sviluppo e la coesione 2007–2013 -formerly FAS”.

\section*{Appendix}
\subsection*{Appendix A}

Let
$$f(L):=-\tanh \left(\sqrt{\frac{\beta}{\mu}}\frac{L}{2}\right) (\lambda L +\alpha) \left(\frac{1}{\sqrt{\beta \mu}}
+\frac{\kappa}{a_0}\sqrt{\frac{\mu}{\beta}}\right)+\frac{\kappa \alpha}{a_0}\frac{L}{2}$$
the left hand side of \eqref{20}.
Notice that
$$
f(0)=0\,, \qquad f'(0)=-\frac{\alpha}{2 \mu} < 0 , \, \forall \;\alpha, \mu > 0.
$$
Moreover,
$$
\lim_{L \rightarrow +\infty}f(L)=\lim_{L \rightarrow +\infty}L\left(-\lambda \left(\frac{1}{\sqrt{\beta \mu}}
+\frac{\kappa}{a_0}\sqrt{\frac{\mu}{\beta}}\right)+ \frac{\kappa \alpha}{2 a_0}\right).
$$
Therefore if $\lambda <\displaystyle{\frac{\alpha \kappa}{2}\frac{\sqrt{\beta \mu}}{(a_0+\kappa \mu)}}$, there exists a positive solution $L$ of \eqref{20}.

\subsection*{Appendix B}
Let
\[
\begin{aligned}
f(V):=&-2 V a_0 - V \frac{\left(a_1 e^{\frac{V}{D}\frac{L}{2}}-a_2 e^{-\frac{V}{D}\frac{L}{2}} \right)}{\sinh\left(\frac{V}{D}\frac{L}{2}\right)}-\frac{\kappa \alpha L}{\tanh\left(\frac{V}{D}\frac{L}{2}\right)}+\\&+2 V^2 \kappa \mu \sqrt{\frac{\mu}{\beta}}\frac{(\lambda L + \alpha)}{(V^2 \mu- \beta D^2)} \frac{\tanh\left(\sqrt{\frac{\beta}{\mu}}\frac{L}{2}\right)}{\tanh \left(\frac{V}{D}\frac{L}{2}\right)}
- \frac{2\kappa D(D^2 \alpha \beta+ V^2 \mu \lambda L)}{V(V^2 \mu- \beta D^2)}
\end{aligned}
\]
the explicit expression of left hand side of \eqref{26}.
Notice that
$$
\lim_{V \rightarrow 0^+} f(V)= \frac{-2D(a_1-a_2)}{L}\,,\quad \lim_{V \rightarrow +\infty} f(V)= -\infty.
$$
Therefore if $a_1<a_2$ there exists a positive solution $V$ of \eqref{26}.

\end{document}